\documentclass[article,twocolumn]{revtex4}
\usepackage{graphicx}
\usepackage{color}
\newcommand{\foonote[2]}{\footnotemark[#1]\footnotetext[#1]{#2}}

\newcommand{\hPi}{\hat{\Pi}}
\newcommand{\hP}{\hat{P}}
\newcommand{\had}{\hat{a}^+}
\newcommand{\ha}{\hat{a}}
\newcommand{\hrho}{\hat{\rho}}
\newcommand{\vac}{|{\rm vac}\rangle}
\newcommand{\ket}[1]{\left| #1 \right\rangle}
\newcommand{\bra}[1]{\left\langle #1 \right|}

\newcommand{\proj}[1]{| #1\rangle\!\langle #1 |}
\newcommand{\projtwo}[2]{| #1\rangle_{#2}\langle #1 |}
\newcommand{\Tr}{\mathrm{Tr}}

\newcommand{\eea}{\end{eqnarray}}
\newcommand{\bea}{\begin{eqnarray}}
\newcommand{\ee}{\end{equation}}
\newcommand{\be}{\begin{equation}}

\begin{document}
\title{Photodetector figures of merit in terms of POVMs}
\author{S.J. van Enk}
\affiliation{Department of Physics and
Oregon Center for Optical, Molecular \& Quantum Sciences\\
University of Oregon, Eugene, OR 97403}
\begin{abstract}
A photodetector may be characterized by various figures of merit such as 
 response time,
 bandwidth,
 dark count rate,
 efficiency, wavelength resolution,
 and photon-number resolution.
 On the other hand, quantum theory says that any measurement device is fully described by its POVM, which stands for Positive-Operator-Valued Measure, and which generalizes the textbook notion of the eigenstates of the appropriate hermitian operator (the ``observable") as measurement outcomes.  Here we show how to define a multitude of photodetector figures of merit in terms of a given POVM. We distinguish classical and quantum figures of merit and issue a conjecture regarding trade-off relations between them. We discuss the relationship between POVM elements and photodetector clicks, and how models of photodetectors may be tested by measuring either POVM elements or figures of merit.
Finally, the POVM is advertised as a platform-independent way of comparing different types of photodetectors, since any such POVM refers to the  Hilbert space of the incoming light, and not to any Hilbert space internal to the detector. 
\end{abstract}
\maketitle
\section{Introduction}
The most general mathematical description of a measurement on a quantum system is in terms of a  positive-operator valued measure (POVM). This is a set---which may be finite or infinite---of hermitian operators $\{\hPi_k\}$ with nonnegative eigenvalues and with $\sum_k \hPi_k=\openone$, where $k$ labels the different measurement outcomes and $\openone$ is the identity operator on the Hilbert space
associated with the system \cite{kraus1983}. The POVM plays a central role in modern quantum information theory. For example, in order to assess how much information an eavesdropper on a quantum cryptographic protocol could have obtained, one has to take into account the most general possible measurement she could have performed \cite{scarani2009}. Furthermore, various different forms of quantum tomography have been developed recently for experimentally determining individual POVM elements $\hPi_k$. Detector tomography \cite{luis1999,feito2009,lundeen2009,zhang2012,natarajan2013,cooper2014,humphreys2015}, self-consistent tomography \cite{mogilevtsev2012}, and SPAM tomography \cite{stark2014,jackson2015,mccormick2017} use different sorts of assumptions to estimate from experimental data what POVM element corresponds to a given outcome of a quantum measurement.

The question considered here is how traditional figures of merit of a photodetector, such as bandwidth, dark-count rate, efficiency, jitter time, response time, spectral sensitivity, and photon-number resolution, are to be expressed in terms of the POVM pertaining to that detector \foonote[1]{Both the POVM and the figures-of-merit may be be functions of external parameters such as operating temperature, bias voltage, etc.}
This may not always be straightforward, as the Hilbert space associated with the radiation part of the electromagnetic field is overwhelmingly large: it is described by infinitely many modes, and each mode in turn is described by an infinite-dimensional Hilbert space. In practice, we often will restrict the Hilbert space of interest, e.g., to some finite wavelength range and some finite time window, but even so the Hilbert space may still be large.

The standard description of the radiation part of the electromagnetic field makes use of four mode numbers, which refer to properties of the classical mode functions, i.e., solutions to the classical source-free Maxwell equations.
For example, the four numbers could describe (i) polarization and the three components of the wave vector [corresponding to the standard expansion of the field in plane waves \cite{cohen}], or (ii) energy, angular momentum, the $z$ component of angular momentum, and parity [corresponding to an expansion in multipole waves \cite{cohen}],
or (iii) energy and the $z$ components of momentum, orbital angular momentum, and spin angular momentum [corresponding to an expansion in Bessel waves \cite{vanenk1994} \foonote[2]{Because of the transversality of radiation fields, there is a subtlety associated with the definition of spin and orbital angular momentum \cite{simmons,cohen}, which, however, is of no concern here.}]. 

The truly nonclassical quantum degree of freedom is 
encoded in the quantum state of the modes. For each mode $i$ (where the label $i$ is a shorthand notation for the four mode numbers) there is an infinite-dimensional Hilbert space which is spanned by the Fock states $\ket{n}_i$ where the nonnegative integer $n$ gives the number of photons in that mode.
In correspondence with this distinction between classical and quantum degrees of freedom, we may divide traditional photodetector characteristics into two groups: one ``classical'' group refers exclusively to physical properties of the classical mode functions of the detected light field. For example, spectral sensitivity, bandwidth, and jitter time all refer to the (classical) spectral degree of freedom, if we consider time as related to the spectrum by the Fourier transform. The second group of photodetector characteristics, which we may call the ``quantum'' group, includes efficiency, dark count rate and photon-number resolution, all of which refer to photon-number statistics, as determined by expectation values of the annihilation and creation operators $\hat{a}_i$ and $\hat{a}_i^+$ for each mode $i$, and products thereof \foonote[3]{The  annihilation and creation operators act on photon-number states as $\hat{a}_i\ket{n}_i=\sqrt{n}\ket{n-1}_i$ and $\hat{a}_i^+\ket{n}_i=\sqrt{n+1}\ket{n+1}_i$.}.

In addition to expressing individual photodetector characteristics in terms of the detector's POVM
we may also wish to derive fundamental tradeoff relations between different characteristics. For example, if we increase the wavelength resolution of our detector, will that necessarily decrease the photon-number resolution?
We may expect tradeoff relations to exist between characteristics that fall within the same group (jitter time and spectral sensitivity being an obvious example, resulting from time-frequency uncertainty relations), but {\em prima facie} not between characteristics that are in the two different groups. For example, photon-number resolution and spectral sensitivity are independent concepts and quantities, and while in practice tradeoff relations may arise from restrictions due to costs or operating temperature or the specific design of the detector, they do not, it seems, arise from fundamental laws of quantum physics.
Indeed, we may write down a mathematically allowed POVM that suffers no such trade offs (see the Discussion and Conclusions section).

Some notational convention used throughout in this paper: $k$ will always be the integer index labeling measurement outcomes and the corresponding POVM elements $\hPi_k$. Modes are indexed by integers $i$, whereas photon numbers are indicated by nonnegative integers $n$. Integers $j$ will be used to label finite-size bins in either the frequency or time domain.
\section{Preliminaries}
\subsection{Generalized measurements}
Standard textbook treatments of quantum measurement talk about an observable $\hat{O}$ as a hermitian or self-adjoint operator, with outcomes represented by the orthogonal set of its eigenvectors (eigenstates), and the corresponding eigenvalues giving the values of the physical quantity measured. But this description represents only a highly idealized subclass  of all possible measurements, the so-called von Neumann measurements.

The modern, fully general, description  of quantum measurements is in terms of a positive-operator valued measure (POVM). Here each outcome is represented by a positive hermitian operator $\hPi_k$ with nonnegative real eigenvalues. (In the special case of an ideal measurement, $\hPi_k$ would be the projector onto the eigenstate of the observable measured, say, $\hPi_k=\proj{k}$, and there is just a single nonzero eigenvalue for $\hPi_k$, namely 1.)
The label $k$ labels the outcome, and the probability of that outcome is determined by the quantum state $\hrho$ of the system on which the measurement is performed, through the Born rule
\be
p_k=\Tr (\hrho \hPi_k).
\ee
The condition $\sum_k p_k=1$ is entailed by completeness $\sum_k \hPi_k=\openone$.
Unlike for ideal measurements, different outcomes $k\neq k'$ do not necessarily correspond to pairwise orthogonal  projectors, and in general we can have
\be
\Tr(\hPi_k\hPi_{k'})\neq 0 \,\,{\rm for} \, k\neq k'.
\ee
This implies, for example, that a repeated measurement does not necessarily repeat the outcome.

Furthermore, measurement outcomes  do not necessarily correspond to projectors onto {\em pure} states, and in general we can have 
\be
\Tr((\hPi_k)^2)<[\Tr(\hPi_k)]^2.
\ee
We may even define a purity for the measurement outcome $k$ as 
\be\label{Pur}
{\rm Pur}(\hPi_k)=\frac{\Tr((\hPi_k)^2)}{[\Tr(\hPi_k)]^2}.
\ee 
This definition is in full analogy to the purity of a quantum state, $\Tr(\hrho^2)$, which becomes even more clear if we define a unit-trace operator $\hrho_k=\hPi_k/(\Tr(\hPi_k)$, so that ${\rm Pur}(\hPi_k)=\Tr(\hrho_k^2)$.
For an ideal von Neumann measurement this purity equals unity, which is the upper bound of the quantity on the right-hand side of (\ref{Pur}). 
The lower bound on the purity is determined by the dimension $d$ of the Hilbert space:
\be
\frac{1}{d}\leq {\rm Pur}(\hPi_k).
\ee
This bound shows that the physical meaning of a POVM element $\hPi_k$ not being pure is that there are multiple orthogonal input states that can lead to the same measurement outcome $k$.
In fact, we may define
an effective (not necessarily integer) Hilbert space dimension  by
\be
d_{{\rm eff}}(k)=\frac{1}{{\rm Pur}(\hPi_k)},
\ee
which then counts how many (at least) such orthogonal states contribute to outcome $k$.

For an example of a POVM in optics, we need look no further than balanced heterodyne detection. The outcome of a heterodyne measurement consists of two real numbers, say  $x$ and $y$, which are combined into a complex number $\alpha=x+iy$. When $x$ and $y$ have been properly normalized, the measurement is represented by the POVM $\{
\frac{1}{\pi}\proj{\alpha}\}$
\foonote[6]{The normalization of the POVM, i.e., the prefactor $1/\pi$,  follows from $\int\! dx\, \int\! dy\,\,\, \proj{x+iy}=\pi
\openone$.}, where the state $\ket{\alpha}$ of the radiation field is a coherent state with amplitude $\alpha$.
Coherent states are pure (and so this POVM is pure), but no coherent state is orthogonal to any other coherent state.
(As this example shows, the effect of interference of external fields with the signal field prior to detecting photo currents is included  in this POVM; similarly, spatial and spectral filtering applied to the signal before counting photons should be included in the POVM description of the detection process as a whole, too \cite{vanenk2017}.)

Another example of a POVM in optics is the measurement of optical phase. Even though no hermitian operator exists that would have all desired canonical properties corresponding to the phase ``observable,'' there is no problem defining the canonical POVM for optical phase \cite{hall1991}.

For examples of how to include in the POVM the effects of standard forms of noise  accompanying the photo detection process, dark counts and finite efficiency, see Refs.~\cite{barnett1998,lee2004,tyc2004,semenov2008,afek2009,audenaert2009,sperling2012,miro2013}.
\subsection{POVM elements and clicks}\label{clicks}
The measurement performed by a photodetector is described fully by its POVM. However, there is no one-to-one correspondence between a single POVM element and a single ``click'' of the detector. Rather,
each POVM element corresponds to one measurement outcome, which refers to the (classical) result at the end of the entire measurement process. Hence, a single measurement outcome may consist of multiple clicks.  For a simple example, suppose our detector has 2 pixels that each can click at most twice  in the short duration the detector is switched on (e.g., because of dead time). 
The total number of different outcomes is then $3^2=9$, since each pixel may click 0,1, or 2 times. Consequently, there are 9 POVM elements in this particular case.

A special measurement outcome is the null outcome, where no clicks are recorded at all. 
We will assume here and in all of the following that our detector is sensitive only to light in a small fraction of all possible modes, and hence that the no-click POVM element  covers the overwhelmingly large part of the full Hilbert space. For that reason it is convenient to label the no-click  outcome with $k={\rm null}$ and write the completeness relation as
\be\label{Pi}
\hPi_{{\rm null}}=\openone-\sum_{k=1}^N \hPi_k=:\openone-\hPi,
\ee
in terms of the $N$ POVM elements $\{\hPi_k,k=1\ldots N\}$ that correspond to nonzero numbers of detector clicks (so in the simple example, $N=8$).
We are going to focus on the information 
obtained from the clicks and  we will discard here the information one might possibly obtain from getting no click.
That is, we assume we are interested in extracting information about photons that are actually present. The null outcome simply and merely means we failed to detect the photon(s). To see why the null outcome carries very little information about what photon(s) may be present, suppose there are very many, say $K\gg 1$ modes not detectable, and a much smaller number, $L\ll K$, that are detectable. Before turning on our detector we lack $\log_2(K+L)$ bits of information about what photon may be present, and after getting the null outcome we have reduced this missing information to (at most) $\log_2(K)$ bits. We thus merely gained (at most) $\log_2(K+L)-\log_2(K)\approx L/K\ll1$ bits.

The operator $\hPi$ defined in (\ref{Pi}) represents all detector clicks together. 
Some of these clicks may be dark counts, not caused by any photon; we do take dark counts into account, see Section \ref{dc}.

\subsection{Discrete modes vs continuum fields}
Since we will only consider the spectral/temporal degree of freedom among the 4 classical degrees of freedom light possesses, in all explicit examples we need just a single mode number, either frequency $\omega$ or time $t$. Both of these are continuous quantities, but Hilbert spaces occurring in quantum mechanics are always taken to be separable, i.e., have a countable basis.
We can go from a continuum description of frequency to a discrete set of modes by a method explained in \cite{blow1990}.
Here is a summary:\foonote[4]{Unlike Ref.~\cite{blow1990} we avoid defining an operator $\had(t)$ here, and do not need to artificially extend the integration over $\omega$ to include negative frequencies.
The (im)possibility of localizing a photon is an interesting theoretical issue in this context, but in practice we may safely leave this can of worms closed; see \cite{keller2005} for an extensive discussion.}

First, we can define a discrete orthonormal set of (complex) mode functions $\{\phi_i(\omega)\}$ normalized such that
\be\label{ortho}
\int_0^\infty\! d\omega\, \phi_i(\omega)\phi_j^*(\omega)=\delta_{ij}.
\ee
Then we can define a creation and annihilation operators for  each discrete mode $i$ by
\bea\label{defa}
\had_i&=&\int_0^\infty\!d\omega\, \phi_i(\omega)\had(\omega),\nonumber\\
\ha_i&=&\int_0^\infty\!d\omega\, \phi_i^*(\omega)\ha(\omega),
\eea
where $\had(\omega)$ and $\ha(\omega)$ are the standard creation and annihilation operators for photons with frequency $\omega$.
The definition (\ref{defa}) is such that the commutation rule $[\ha_i,\had_{i'}]=\delta_{ii'}$ follows from
$[\ha(\omega),\had(\omega')]=\delta(\omega-\omega')$ and Eq.~(\ref{ortho}) \foonote[5]{If in addition we assume completeness in $L^2[0,\infty]$ of the set of functions $\{\phi_i(\omega)\}$, then we can expand
$\had(\omega)=\sum_i \phi_i^*(\omega)\had_i$.}. The integration range here and in Eq.~(\ref{ortho})  extends over just the positive frequencies, since creation operators $\had(\omega)$ are defined only for those frequencies. 
A pure single-photon state containing exactly one photon in mode $i$ is defined as
\be\label{aom}
\ket{\phi_i}=\int_0^\infty\! d\omega \,\phi_i(\omega)\had(\omega)\vac=\had_i\vac,
\ee
with $\vac$ the vacuum state containing no photons.
States of multiple photons can be described in terms of the operator $\had_i$, too. For example, the state with exactly $n$ photons in mode $i$ (and none elsewhere) is
\be
\ket{n}_i=\frac{(\had_i)^n}{\sqrt{n!}}\vac,
\ee
and a coherent state of mode $i$ with complex amplitude $\alpha$ is
\bea
\ket{\alpha}_i&=&
\exp(\alpha\had_i-\alpha^*\ha_i)\vac\nonumber\\
&=&\exp(-|\alpha|^2/2)\sum_n
\frac{\alpha^n}{\sqrt{n!}}\ket{n}_i.
\eea

For an {\em ideal} 100\% efficient detector that detects the presence of photons (not their energy), the probability $dP_i(t)$ to detect a photon  in the state $\ket{\phi_i}$ during a small  time interval between $t$ and $t+dt$ would be 
\be\label{Pit}
dP_i(t)=\frac{dt}{2\pi}\left|\int_0^\infty \!d\omega\,\phi_i(\omega)\exp(-i\omega t)\right|^2,
\ee
which is such that the ideal single-photon detection rate, $P_i(t)$, integrates to 1:
\be
\int_{-\infty}^\infty\!dt\, P_i(t)=\int_0^\infty\! d\omega\, |\phi_i(\omega)|^2=1.
\ee
\section{Detector characteristics in terms of POVMs}
In the first two subsections to follow we will discuss the information gained about a single photon from a single click of the detector.  Subsequent subsections discuss what sort of information is obtained about the presence of multiple photons and what information is obtained from multiple clicks.
\subsection{Single-photon bandwidth}
First let us define the single-photon part of an arbitrary POVM element,
\be\label{Pione}
\hPi^{(1)}_k=\hP^{(1)}\hPi_k \hP^{(1)}.
\ee
Here $\hP^{(1)}$ is the projector onto the (relevant part of) the 1-photon subspace, defined by a sum over all modes $i$ (with $\ket{\phi_i}$ defined in (\ref{aom}))
\be
\hP^{(1)}=\sum_i \proj{\phi_i}.
\ee
We can now define  two bandwidth-related quantities.
First,
\be\label{band1}
\Omega_k^{(1)}=\Tr(\hPi_k^{(1)})
\ee
is the effective size of the single-photon Hilbert space covered by outcome $k$. For example, suppose outcome $k$ represents the click of one particular pixel $k$ that is sensitive only to one particular wavelength. If that pixel detects a photon with that  wavelength with probability $p<1$, then 
we would have $\Omega_k^{(1)}=p<1$.
Normally, one expects sensitivity to a range of wavelengths such that $\Omega_k^{(1)}>1$ or even $\Omega_k^{(1)}\gg1$.
The second quantity we define is
\be
\Omega^{(1)}=\sum_{k=1}^N \Omega^{(1)}_k=\Tr(\hP^{(1)}\hPi\hP^{(1)}),
\ee
which is a measure of the effective size of the single-photon Hilbert space covered by all of the detector's possible clicks. Note that these definitions of bandwidth are all basis-independent and dimensionless, and in particular do not distinguish between spectral and temporal degrees of freedom. The bandwidth thus defined is appropriate in a communication context, in which $\Omega^{(1)}$ would roughly  be the total number of single-photon channels that could be detected (but not necessarily distinguished).

For an arbitrary single-photon POVM element we can find its diagonal form
\be\label{diagk}
\hPi_k^{(1)}=\sum_i w_i^{(k)} \proj{\phi_i^{(k)}},
\ee
where $\ket{\phi_i^{(k)}}$ denotes a pure single-photon state of mode $i$ (see Eq.~(\ref{aom})).
Note that each POVM element $\hPi_k^{(1)}$ may be diagonal in a different basis $\{\ket{\phi_i^{(k)}}\}$, hence the superscript $(k)$.
The weight $w_i^{(k)}$ has the meaning of the conditional probability of getting measurement outcome $k$ given an input state $\ket{\phi_i^{(k)}}$,
\be
\Pr(k|i)=\Tr(\hPi_k^{(1)}\proj{\phi_i^{(k)}})=
\bra{\phi_i^{(k)}}\hPi_k^{(1)}\ket{\phi_i^{(k)}}=w_i^{(k)},
\ee
and so it
necessarily lies between 0 and 1. Moreover, their sum over $i$ gives
the bandwidth (\ref{band1}): $\sum_i w_i^{(k)}=\Omega^{(1)}_k$.

What does, conversely, the measurement outcome $k$ imply about the input state of the photon that was just detected? Suppose the input states
$\ket{\phi_i^{(k)}}$ for different $i$ appear {\em a priori} with some probability $\Pr(i)$.
After getting outcome $k$ we can update our probability distribution over input states $i$ to
\be
\Pr(i|k)=\frac{w_i^{(k)}\Pr(i)}{\Pr(k)},
\ee
with $\Pr(k)=\sum_i \Pr(i)w_i^{(k)}$ the {\em a priori} probability to get outcome $k$.
We can quantify the amount of information we still lack about which mode the photon was in by the Shannon entropy
\be
H^{(k)}=-\sum_i \Pr(i|k)\log_2 \Pr(i|k).
\ee
This quantity depends on our prior knowledge (or prior assumptions) about the 
input states. We can eliminate this dependence by making $\Pr(i)/\Pr(k)$ independent of $i$.
In that case we would have 
$
\Pr(i|k)=w_i^{(k)}/\Omega_k^{(1)},
$
and the corresponding Shannon entropy is then an effectively input-independent  quantity---and we'll use a calligraphic script to emphasize this useful property---that characterizes the detector,
\be
{\cal H}^{(k)}=-\sum_i \frac{w_i^{(k)}}{\Omega_k^{(1)}}\log_2 \frac{w_i^{(k)}}{\Omega_k^{(1)}}.
\ee
In terms of $\hrho_k=\hPi_k^{(1)}/\Tr(\hPi_k^{(1)})$, we get the manifestly input-independent
\be
{\cal H}^{(k)}=-\Tr[\hrho_k\log_2\hrho_k].
\ee
We could also use the so-called collision entropy (which is the Renyi entropy $H_{\alpha}$ of order $\alpha=2$) to quantify our lack of knowledge as
\be
H_{\alpha=2}^{(k)}=-\log_2 (\sum_i \Pr(i|k)^2).
\ee
Again, when we assume $\Pr(i)/\Pr(k)$ is independent of $i$, this collision entropy becomes input-independent, and in fact we get
\be
{\cal H}_{\alpha=2}^{(k)}=-\log_2
({\rm Pur}(\hPi_k^{(1)})),
\ee
with the purity
Pur$(.)$  defined in (\ref{Pur}).

And so the purity of $\hPi_k^{(1)}$ and the Shannon entropy quantify in different but well-defined ways the lack of specificity of the outcome $k$.
In the following subsections we use these same ideas to define spectral and timing resolution, as well as the photon-number resolving capabilities of a given detector. We choose to utilize the Shannon entropy there, but could use the collision entropy or the purity just as well.
\subsection{Wavelength and timing resolution}\label{wt}
Given the diagonal form (\ref{diagk}) of $\hPi_k^{(1)}$  we find the normalized {\em a posteriori} probability distribution over $\omega$ that outcome $k$ implies as
\be
\Pr(\omega|k)=\sum_i\Pr(i|k)|\phi_i^{(k)}(\omega)|^2.
\ee
Analogously, we can define an {\em a posteriori}  probability distribution over detection times of the photon
as
\be
\Pr(t|k)dt=\sum_i\Pr(i|k)dP_i(t),
\ee
with $dP_i(t)$ defined in (\ref{Pit}).
These probability distributions are over continuous quantities. In practice finite precision forces one to bin the frequency and time measurements into finite-sized intervals.
So, let us first divide the frequency range into equal-sized \foonote[7]{We could drop the assumption of equal-sized frequency intervals and switch to, say, equal-sized intervals in wavelength.} small frequency intervals $\delta \omega$. Then, given the probability distributions $\Pr(\omega|k)$ for each POVM element $\hPi_k^{(1)}$, we may define for each positive integer $j$ the probability
\be
p(j|k)=\int_{(j-1)\delta \omega}^{j\delta\omega} d\omega\, \Pr(\omega|k),
\ee
which is the {\em a posteriori} probability for the detected photon to belong to frequency bin $j$.
The Shannon entropy
\be
H_\omega^{(k)}=-\sum_j p(j|k)\log_2p(j|k),
\ee
properly quantifies the amount of information (in units of bits) we still lack after having obtained outcome $k$ about which frequency interval the photon we just detected belongs to. 

We can define the analogous probabilities $q(j|k)$ for integers $j$ (not necessarily positive) and the corresponding entropy $H_t^{(k)}$ for the lack of information about the photon's time of detection once we have divided time in bins of finite size $\delta t$, as 
\bea
q(j|k)&=&\int_{(j-1)\delta t}^{j\delta t} dt\, \Pr(t|k),\nonumber\\
H_t^{(k)}&=&-\sum_j q(j|k)\log_2q(j|k).
\eea
For a given POVM we can take the weighted averages of the individual entropies
\be
\bar{H}_{\omega,t}=\sum_{k=1}^N \frac{\Omega_k}{\Omega}H_{\omega,t}^{(k)},
\ee
as a measure of how much information about frequency or time
we anticipate lacking on average. 

For each outcome $k$, we have the entropic uncertainty relation \cite{bialynicki1975} \foonote[8]{Uncertainty relations for the collision entropy, even including binning, as well as other Renyi entropies, can be found in  \cite{bialynicki2006}.}
\be\label{unc}
H_\omega^{(k)}+H_t^{(k)}>\log_2(e)-1-\log_2\frac{\delta\omega\delta t}{2\pi}.
\ee
The two weighted averages satisfy the same uncertainty relation, since the right-hand side of (\ref{unc}) is independent of $k$, i.e., we also have 
\be
\bar{H}_\omega+\bar{H}_t>\log_2(e)-1-\log_2\frac{\delta\omega\delta t}{2\pi}.
\ee
The entropies we have defined here do depend on our choices of $\delta\omega$ and $\delta t$.
The smaller we pick our bin sizes, the larger will be the missing information.
Roughly speaking, each time we make the interval smaller by a factor of 2, $\bar{H}$ increases by approximately 1 bit (in fact, by at most 1 bit).
This strong dependence on bin size is not a desirable property, even though we can still make sensible comparisons between different detectors for given values of $\delta\omega$ and $\delta t$. To obtain a more useful (and dimensionful) quantity we
 could adopt the following convention. Pick interval sizes $\delta \omega$ and $\delta t$ such that the averaged missing information $\bar{H}_{\omega,t}$ equals a few bits. That is, these intervals are really too small to be resolved by the detector. 
If we define
\bea
\Delta\omega&=&2^{\bar{H}_{\omega}} \delta\omega\nonumber\\
\Delta t&=&2^{\bar{H}_{t}} \delta t
\eea
then the quantities on the left-hand side, $\Delta\omega$ and $\Delta t$, satisfy
an uncertainty relation independent of the bin sizes $\delta\omega$ and $\delta t$,
\be
\Delta\omega \Delta t\geq e\pi\approx 8.54.
\ee
We could take $\Delta\omega$ and $\Delta t$ as measures of the average frequency and timing resolutions of our detector, respectively, even though they still weakly depend on the bin sizes. (We could simply require the entropies to equal, say, 4 bits, in order to fix the bin sizes and make the definitions for $\Delta\omega$ and $\Delta t$ unique.)
 \subsection{Photon number resolution}
In order to talk about photon-number resolution we do need to first distinguish  between different modes. 
Let us fix one mode of interest, $i$, so that the Hilbert space of interest is spanned by the Fock states $\ket{n}_i$.
Then, for a given outcome $k$ we need the {\em a posteriori} probability distribution over different numbers of photons in mode $i$ that outcome $k$ implies. We may write this {\em a posteriori} probability as a conditional probability $\Pr(n|k,i)$: given outcome $k$ and given an input mode $i$, what is the probability for a number of photons equal to $n$?
The entropy
\be
H_{n,i}^{(k)}=-\sum_n \Pr(n|k;i)\log_2\Pr(n|k;i)
\ee
quantifies (in bits) how much information concerning the photon number $n$ in mode $i$ is still missing after we have obtained outcome $k$.

Again we could assume, for the purpose of an input-independent definition, that the {\em a priori} probabilities of different numbers  of photons are equal over some finite range. If we define weights
\be
\Omega_{k,i}^{(n)}= \bra{{\rm vac}}(\ha_i)^n\hPi_k(\had_i)^n\vac/n!,
\ee 
 the sought-after {\em a posteriori} probability distribution over $n$ is then given by
\be
\Pr(n|k,i)=\frac{\Omega_{k,i}^{(n)}}{W_{k,i}},
\ee
where 
$
W_{k,i}=\sum_{n} \Omega_{k,i}^{(n)}.
$ The Shannon entropy may be written then as
\be
{\cal H}_{n,i}^{(k)}=-\sum_n \frac{\Omega_{k,i}^{(n)}}{W_{k,i}}\log_2\frac{\Omega_{k,i}^{(n)}}{W_{k,i}}.
\ee
(And again we may average this quantity over all POVM elements by summing either over $k$ or over $i$ or over both, and giving each term a relative weight $W_{k,i}$.)

The sums over $n$ here all extend in principle over the entire range of allowed $n$ but in practice these sums really contain only a few non-neglible terms. For instance, if we have an array of highly efficient on/off detectors and two of them click, then the probability that more than a dozen of photons (in the right wavelength range) caused just these two clicks  is negligible. 
 
It is really simpler here to use the purity to quantity photon-number resolution. We first restrict the POVM element $\hPi_k$ to mode $i$,
\be
\hPi_k^{(i)}=\sum_n \projtwo{n}{i}\hPi_k \projtwo{n}{i},
\ee 
and then get the collision entropy as
\be
{\cal H}_k^{(i)}=-\log_2 {\rm Pur}(\hPi_k^{(i)}),
\ee
as a measure for how specific outcome $k$ 
is about the number of photons in mode $i$ that caused it.

Note, finally, that there is an entropic uncertainty relation for photon number and phase like that between time and frequency.
However, phase sensitivity is usually not considered a property of the detector, but rather of the interferometric setup in which the photodetector is placed. That is why we will not consider phase sensitivity  here.
\subsection{Efficiency and dark count rates}\label{dc}
Efficiency is defined as the probability that a single photon (in a given mode) is detected. Clearly, the efficiency is in general a mode-dependent quantity.
We can diagonalize not just each individual single-photon detection POVM element (as we did in Eq.~(\ref{diagk})), but their sum 
\be
\sum_{k=1}^N \hPi_k^{(1)}=\hP^{(1)}\hPi\hP^{(1)}=\sum_i w_i \proj{\phi_i}.
\ee
The weights $w_i$ appearing here are really efficiencies for the modes $i$: $w_i$ is the probability that a single photon present in mode $i$ causes a click. And so we can simply define efficiencies $\eta_i=w_i$. The largest of all $w_i$s is the efficiency of the detector at its most sensitive point,
\be
\eta_{\max}=\max_i w_i.
\ee
If we are interested in a particular mode $i'$ that is {\em not} a basis vector in the basis that diagonalizes  $\sum_{k=1}^N \hPi_k^{(1)}$, then we can still  define the appropriate single-photon detection efficiency as
\be
\eta_{i'}= \bra{\phi_{i'}}\hPi \ket{\phi_{i'}}=
\bra{{\rm vac}} a_{i'}\hPi a_{i'}^+\vac.
\ee
Dark count rates are determined by the probabilities of detector clicks when no photon is present. So, we clearly need the quantities
\be
d_k=\Tr (\hP^{(0)}\hPi_k \hP^{(0)})=\bra{{\rm vac}}
\hPi_k\ket{{\rm vac}}=\Omega_k^{(0)}.
\ee
The dark-count rate is not mode dependent, but it may depend on $k$.
The dark count rate for a detector that is switched on for a duration $T$ is in fact
\be
d=\frac{\sum_{k=1}^N d_kN(k)}{T},
\ee
because the numerator equals the expected number of dark counts provided we let $N(k)$ denote the total number of clicks occurring in outcome $k$ (recall Section \ref{clicks}).

\subsection{Response time and detection rate}
Maximum rate and dead time or response time are determined by correlations in time between multiple clicks.  The response time may depend on the mode detected. 
Consider a single-photon state of the form
\be
\ket{\phi}=\int_{0}^\infty\,d\omega\, \phi(\omega)\had(\omega)\vac
\ee 
and consider a time translated version of this state (i.e., the state that would result from free evolution of $\ket{\phi}$ over some time $\tau>0$)
\be
\ket{\phi_\tau}=\hat{T}(\tau)\ket{\phi}=:
\int_{0}^\infty\,d\omega\, \phi(\omega)\exp(-i\omega \tau)\had(\omega)\vac.
\ee
If $\tau$ is not much larger than the response time we expect this quantity:
\be
P(0,\tau):=\bra{\phi,\phi_\tau}\hPi\ket{\phi,\phi_\tau},
\ee
i.e., the joint probability to detect both a photon in mode $\phi$ and one in the time-translated mode $\phi_\tau$, to be less than the product of the two individual single-photon detection probabilities
\bea
P(0)&:=&\bra{\phi}\hPi\ket{\phi},\nonumber\\
P(\tau)&:=&\bra{\phi_\tau}\hPi\ket{\phi_\tau}.
\eea
If, for a given mode, we define the response time as the time it takes to go from 10\%  to 90\% of maximum detection probability (which we assume is associated with detection at $\tau=0$), then we need the time delays $\tau_{10}$ and $\tau_{90}$ such that
\bea
P(0,\tau_{10})&=&\frac{1}{10}P(0)^2,\nonumber\\
P(0,\tau_{90})&=&\frac{9}{10} P(0)^2.
\eea
This assumes $\tau_{90}>\tau_{10}$, and if there is no such $\tau_{90}$ satisfying the above requirement for a given $\tau_{10}$, then we can set $\tau_{90}=\infty$.
The response time for a particular mode $\phi_i$ is then defined as
$\theta_i=\tau_{90}-\tau_{10}$.
A total detection rate can then be defined as a sum of inverse response times $\theta_i^{-1}$ for the modes $i$ that diagonalize $\sum_{k=1}^N\hPi_k^{(1)}$,
\be
R=\sum_i \theta_i^{-1}.
\ee
This quantity automatically takes into account the possibility of having a large array of parallel detectors. Even if each of the detectors in the array has a slow response, the rate $R$ may still be high.
\section{Discussion and conclusions}
We have shown here how standard photo detector figures of merit can be directly expressed in terms of the POVM describing the quantum properties of the photo detection. Since the POVM is fully quantum-mechanical, so are the figures of merit thus defined.

One advantage shared by the standard figures of merit and the POVM 
is that they do not refer to the Hilbert spaces internal to the photodetector (the Hilbert spaces associated with phonons, excitons, polaritons, discrete energy levels of single absorbers, etc., etc.), but only to the Hilbert space and properties of the photons that are being detected. 
The advantage of the POVM over the figures of merit is that it, in principle, contains {\em all} information about how the photodetector's clicks provide information about the incoming photons.

 A quantum field theory description of light shows that the standard detector characteristics fall into two groups: one group refers to the classical degrees of freedom of the classical mode functions, the other to the quantum degree of freedom related to photon statistics.
The conjecture is that there are no {\em fundamental}---as opposed to practical--- tradeoff relations between characteristics from the different groups. To illustrate this, consider the following POVM, perfectly legitimate from the mathematical point of view. For a given set of orthogonal modes $\{\phi_i(\omega)\}$, define
\be
\hPi_n^i=\ket{n}_i\!\bra{n},
\ee
with $\ket{n}_i$ the state of exactly $n$ photons in mode $i$ (and no photons in any other mode).
Every mode $i$ must satisfy a time-frequency uncertainty relation for its mode function (and we gave  such relations in two different forms in Section \ref{wt}), but the measurement is perfectly number-resolving, and has zero dark counts irrespective of the choice of basis $\{\phi_i(\omega)\}$. 

Photon-number resolution and spectral and timing resolution were all defined here in terms of entropic quantities. The latter 
quantify the amount of information still missing (about photon number, wavelength, and time of arrival, respectively, of the input light) after we have obtained a particular measurement outcome. These entropic quantities are all dimensionless, but we also showed how dimensionful quantities like bandwidth (in Hz) or timing resolution (in seconds) may be obtained from the entropic quantities.

Finally, the purity of a POVM element seems a useful additional (nontraditional) figure-of-merit: roughly speaking, it quantifies how many different orthogonal quantum input states could lead to exactly the same measurement outcome.
 
The strategy followed in this paper was to consider the POVM as given. The next questions to be considered are (i) how one obtains such a POVM description, and (ii) how to experimentally test it. 
The photo-detection problem in all its generality is too complicated to allow for an {\em ab initio} solution, and one will have to
resort to simplified physical models that can, at a minimum, be used to fit to data. Model selection \cite{burnham2003} is then a nice statistical technique that allows one to rank different models,  based on how well the models fit the data and how many fitting parameters they use.
This technique is especially useful for reducing the number of parameters if the 
relevant Hilbert space is large \cite{schwarz2013,ferrie2014}, as it indeed is for the photo-detection problem. 

One way to get relevant test data is to perform small-scale detector tomography (large-scale tomography is not feasible). That is, by restricting oneself to a not-too-large Hilbert space (spanned by, say, the states with at most, say, 20 photons \cite{feito2009} in 1 or 2 modes, as detected by one particular pixel), one may experimentally estimate the corresponding POVM elements. These estimates can be used directly to evaluate one's models.
While tradeoff relations obtained within such a {\em model} may not constitute the {\em fundamental} limits of photodetection, they should be of great practical interest nonetheless. 

The other way to test model descriptions, is by measuring 
figures-of-merit like quantum efficiency as a function of, say, wavelength (at different temperatures or while varying other control parameters) and comparing the result to what the underlying theory says about this functional dependence (either directly or indirectly via the POVM). This approach has been successfully adopted in several recent experiments on nanowire superconducting single-photon detectors \cite{renema2012,renema2013,renema2014,wang2015,gaudio2016}.
\section*{Acknowledgements} 

I thank  Michael Raymer and Andrzej Veitia, as well as the participants in the DARPA DSO Detect Theory Kickoff and Technical Exchange Meeting for useful discussions.

This work is supported by funding from
DARPA under
Contract No. W911NF-17-1-0267.
\bibliography{photo_detection_new}
\end{document}